# Structural design of the acrylic vessel for the Jinping Neutrino Experiment

Zongyi Wang[a], Yuhao Liu[a,*], Shaomin Chen[b], Yuanqing Wang[c], Zhe Wang[b], Ming Huang[a]

a. School of Civil and Hydraulic Engineering, Huazhong University of Science and Technology, Wuhan 430074, China

b. Department of Engineering Physics, Tsinghua University, Beijing 100084, China

c. Department of Civil Engineering, Tsinghua University, Beijing 100084, China

*Corresponding author: Yuhao Liu. E-mail address: m202371796@hust.edu.cn

**Abstract:** The Jinping neutrino experiment is designed to have multiple purposes in the China Jinping Underground Laboratory. Following the acrylic vessel design requirements proposal, a structural scheme has been developed and optimized. Subsequently, the stability of the acrylic shell structure was calculated using finite element analysis, as well as the load-bearing capacities under various working conditions. Further, the effects of temperature changes, rope failures, and Young's modulus of the ropes on the static behavior of the structure were analyzed. The results indicated that the stress level and structural displacement of the structure scheme satisfies the design requirements, as well as the stability of the vessel under compression. Moreover, the stress and displacement of the acrylic shell satisfies the given working conditions and temperatures. The structural scheme ensures basic safety if the rope fails.

## 1. Introduction

Neutrinos are among the most enigmatic particles yet to be fully understood, leading to astonishing breakthroughs [1]. To date, several Nobel prizes in physics have been awarded in neutrino research, underscoring the substantial impact of neutrino studies on the human quest to understand the essence of the universe [1-2].

According to the planetary synthesis theory, the sun may produce an abundance of neutrinos via nuclear fusion [3]. These neutrinos can serve as independent probes for solar energy generation mechanisms. The primary scientific goal of the Jinping Neutrino Experiment (JNE) is to observe solar neutrinos to understand the thermonuclear reactions of the sun and investigate the dynamics inside massive stars [4-6].

Neutrino detectors are required for the detection of neutrinos. The core vessel of some well-known neutrino detectors in the world are made of acrylic and their structures can be categorized into two types: cylindrical structure, and spherical structure. The Daya Bay [7], RENO [8], and Double Chooz [9] are classic cylindrical structures. The Daya Bay detector consists of two acrylic vessels. The inner acrylic vessel is about 3 m in diameter and 3 m in height, and the outer one is approximately 4 m in diameter and 4 m in height. The core detectors of the SNO+ [10] and JUNO [11-13] are two neutrino detectors with spherical acrylic vessels. The SNO+ detector contains an acrylic shell with a diameter of 12 m, which is supported through synthetic fiber ropes. This detector can hold 20 kton of liquid scintillator. The JUNO detector has a huge acrylic shell (about 35.4 m in diameter) supported by a stainless-steel lattice shell through steel structure components. The core vessels of some other neutrino detectors were made of nylon, rather than acrylic, such as Borexino [14-15] and KamlanD [16]. The Borexino detector was made of an 8.5-diameter nylon sphere with 300 ton of scintillator, and the KamLAND detector contains a 13 m diameter balloon made of nylon 6 filled with nearly 1 kton of liquid scintillator.

The JNE detector will be installed at the China Jinping Underground Laboratory (CJPL), nestled beneath Jinping Mountain in the southwest. The CJPL has emerged as the deepest underground laboratory in the world and has a vertical rock cover of roughly 2400 m (Fig. 1). Several relevant studies [17-19] have been conducted to ensure that the environmental background fulfills the needs of the Jinping solar neutrino experiment.

In a pioneering effort, Wang et al. developed a one-ton prototype for the JNE [20]. They selected an acrylic shell structure as the central vessel type for the JNE detector. The prototype featured an acrylic shell with a 1.29 m inner diameter and 20 mm thick, supported mainly by a stainless-steel truss and further supported by an acrylic bearing at the base. Wang et al. examined the mechanical attributes of an acrylic container and served as a reference for the design of an actual JNE detector.

The mechanical structure of the JNE detector consists of several key components, including acrylic shell, steel frames and synthetic fiber ropes.

This paper explores the design and analysis of acrylic shell and fiber ropes in the main structure of the detector. In Section 2, the performance requirements and design characteristics of the detector's core vessel are discussed, followed by an introduction to various working conditions. In Section 3, an acrylic shell and its supporting system was designed; the scheme was optimized by analyzing the rope's position; the accurate stress distribution and buckling safety factor of the acrylic shell were obtained by finite element analysis (FEA); the load-bearing capacities of the optimized scheme was calculated under various working conditions to ensure the structural safety. In Section 4, the study further conducted a series of analyses, including the temperature effects, rope failure, and rope Young's modulus, to ascertain the safety and reliability of the proposed design.

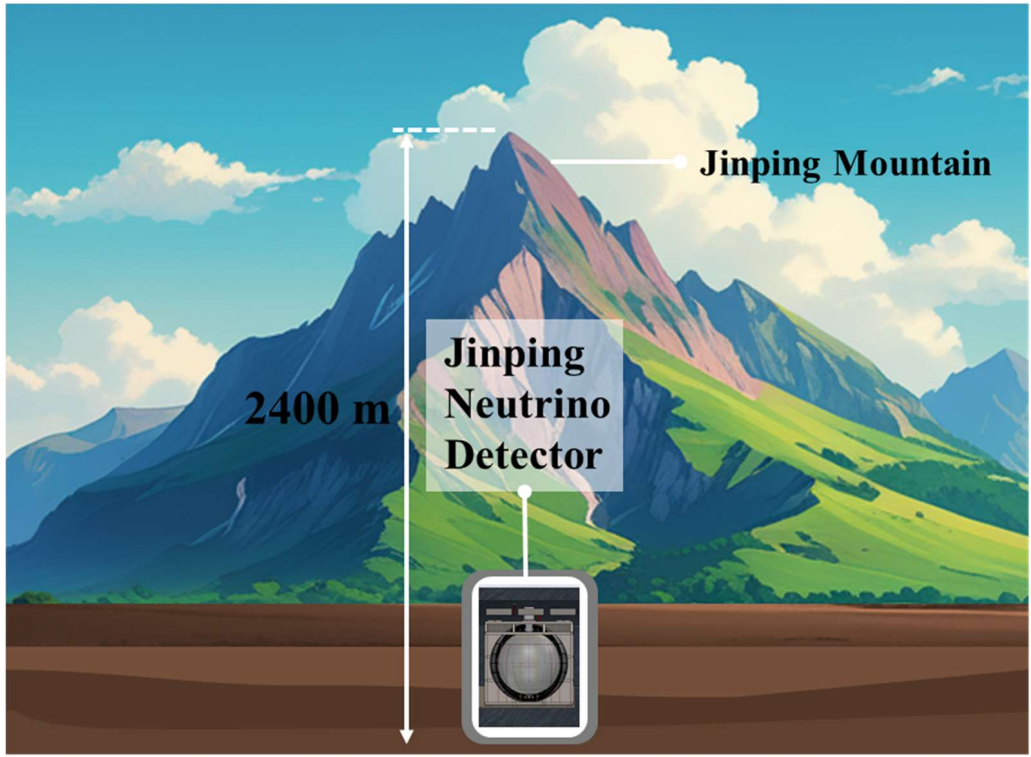

Fig. 1 Location of the JNE detector

## 2 Design Consideration of the JNE acrylic vessel

### 2.1 Working Conditions

The JNE detector will be a hybrid detector [4-6] with a capacity of 500 m$^3$. Its interior and exterior will be filled with purified water (density 1.0 g/cm$^3$), liquid scintillator (assumed to be 0.8 g/cm$^3$ [21]), or even more types of target materials, including a novel type of slow scintillating liquid (density 1.2 g/cm$^3$) that Guo et al. have proposed for using in the JPE [22-23]. This means that under normal working conditions, the acrylic shell will face the situation in Fig. 2, where the liquid densities inside and outside the acrylic shell are unequal. When the liquid is replaced, the liquid levels inside and outside the shell are not similar. Table 1 lists the conditions considered. Considering the fluid exchange rate of the JNE detector, a maximum liquid-level difference of 2 m was sufficient. Here, $h_1$ is the outer liquid level, and $h_2$ is the inner liquid level, as shown in Fig. 2. Synthetic fiber ropes were used to restrain and support the acrylic shell, but not shown in Fig. 2.

**Table 1 Working conditions**

| Working condition | Density of acrylic (g/cm$^3$) | Density of inner liquid (g/cm$^3$) | Density of outer liquid (g/cm$^3$) | Inner liquid level $h_1$ (m) | Outer liquid level $h_2$ (m) | Description |
|---|---|---|---|---|---|---|
| 1 | | No liquid inside or outside the acrylic vessel | | | | Short-term condition |
| 2 | | | | 14 | 14 | Long-term condition |
| 3 | | 1.2 | 1.0 | 12 | 14 | Short-term condition |
| 4 | | | | 14 | 12 | Short-term condition |
| 5 | 1.19 | | | 14 | 14 | Long-term condition |
| 6 | | 1.0 | 1.2 | 12 | 14 | Short-term condition |
| 7 | | | | 14 | 12 | Short-term condition |
| 8 | | 0.8 | 1.0 | 14 | 14 | Long-term condition |
| 9 | | 1.0 | 0.8 | 14 | 14 | Long-term condition |

Note: (a) When the density of inner liquid is 0.8 (or 1.0) g/cm$^3$ and that of the outer liquid is 1.0 (or 0.8) g/cm$^3$, the conditions considering liquid level differences are not considered for brevity. (b) The stresses on acrylic under long-term working conditions should be lower than 3.5 MPa, and the stresses under short-term working conditions should be limited within 7.0 MPa.

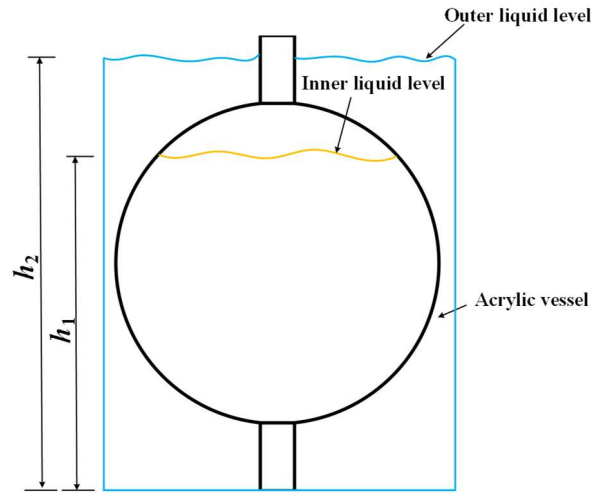

**Fig. 2 Liquid levels insider and outside the acrylic vessel**

## 2.2 Basic Requirements

The material used for the acrylic shell of the JNE detector is polymethyl methacrylate (PMMA), a classic viscoelastic material [24]. Owing to its high strength, exceptional light-transmission properties, and low background radiation, it is frequently selected as a material for neutrino detectors [25]. Its transparency can easily exceed 90%, and its tensile strength can reach 50–90 MPa [26-27]. However, considering the large size and future demands of the JNE detector, there are specific requirements for structural stress and displacement to ensure long-term reliability and stability. In addition, the mechanical properties of fiber ropes, which are the primary support materials for acrylic shells, must be considered.

To satisfy the bearing capacities and safety performance of the JNE detector structure, the following requirements must be considered before proceeding with the design and calculations:

1) Stress limitations of the acrylic. The von Mises stress, formulated from the perspective of the fourth strength theory, evaluates the internal stress distribution in materials under multiaxial stress. Moreover, for brittle materials, such as PMMA, brittle failure is primarily caused by tensile stress, which is related to the first strength theory measured by the first principal stress. This stress represents the maximum tensile stress of the material in a particular direction. In practical applications, both theories are expected to be considered. Stachiw suggested that the conservative allowable working stress of acrylic materials is 10.335 MPa [26]. However, PMMA exhibits an apparent creep effect as a classic viscoelastic material. Creep refers to the phenomenon in which deformation increases gradually with time under the action of a constant force. Additionally, flaws or microcracks significantly affect the mechanical properties of PMMA [28]. Therefore, the stress levels of the acrylic structure must be maintained at low levels. In view of the importance of the JNE detector, this study set a long-term stress limit of 3.5 MPa to control creep in the acrylic shell and a short-term stress limit of 7.0 MPa is recommended for short-term and extreme conditions, such as rope failure, to prevent acrylic fractures.

2) Safety factor limitation considering buckling instability of the acrylic vessel. Buckling instability of a structure refers to the phenomenon in which an irreversibly large deformation occurs after the compressive load exceeds a specific limit. This phenomenon can lead to significant brittle failure of the acrylic shell [25]. The safety factor is an essential measure for evaluating the buckling performance of a structure; a higher safety factor indicates more excellent stability under the given conditions. In the Jinping experiment, the differential liquid densities inside and outside the acrylic shell imposed substantial pressure, increasing the possibility of buckling instability. In the absence of specific codes for acrylic spherical pressure vessels [29-31] and considering the standards for steel pressure vessels, it is prudent to target a safety factor exceeding 4.0 in the context of geometrically nonlinear analysis.

3) Displacement limitation of the acrylic vessel. The structural components of the JNE detector also include a steel frame, and PMTs, as shown in Fig. 3. A distance of 900 mm was maintained between the acrylic shell and the steel truss. However, PMTs, which are 300 mm in height and are used for detecting Cherenkov radiation, must be installed between the acrylic shell and the steel truss, leaving only 600 mm of movement space for the acrylic shell. The acrylic shell was subjected to floatage or gravity effect under different working conditions as shown in Fig. 3, and therefore the shell would move upward or downward. It was specified that the displacement of the acrylic shell should not exceed 600 mm under any conditions.

4) Axial force limitation of the ropes. The acrylic shell of the JNE detector is supported by synthetic fiber ropes, as shown in Fig. 3; thus, the safety of the acrylic shell is significantly linked to these ropes. It is essential to prevent ropes from breaking or elongating greatly, as such failures could precipitate the collision of the acrylic shell with the stainless-steel truss or PMTs, leading to structural damage. Hence, it is recommended that the axial force on synthetic fiber ropes should be less than 15% of their breaking force, ensuring that the ropes will not quickly fail owing to creep or breaking, and the breaking axial force of the ropes is assumed to be 1000 kN in this study.

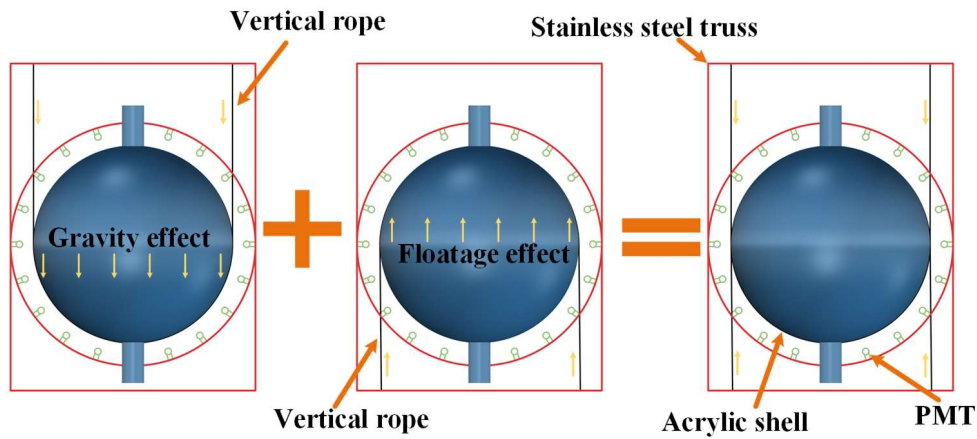

Fig. 3 Schematic diagram of the JNE Detector

## 2.3 Design Requirements

The JNE detector is expected to perform better and more conveniently than previous detectors. The acrylic shell should be designed to satisfy the following structural and functional requirements as effectively as possible.

1) Easy to replace liquid. Currently, most neutrino detectors worldwide use single-target materials, such as pure water or various liquid scintillators, with only one pipeline, which cannot interchange the target and shielding materials. The primary characteristic of the JNE detector's structure is its ability to quickly replace various target materials, thereby enabling multiphase detection technology to conduct experiments with different target and shielding materials. The JNE detector can observe solar neutrinos and perform versatile experiments, such as exploring geoneutrinos in the Himalayan region and conducting neutrino-less double-beta decay experiments.

2) Ensuring structural safety during single-rope failures. Because the support system of an acrylic shell consists of synthetic fiber ropes, which differ from conventional steel supports, the mechanical properties of these ropes can be easily affected by external factors such as chemical corrosion and friction fracture. The destruction of steel is ductile, and there will be significant deformation before failure; hence, it gives sufficient warning to the problem and remedies it in time. Fiber rope damage is brittle damage that causes people to be unprepared. Relying solely on stress limitations to avert rope breakage is insufficient, as countless unforeseen circumstances may lead to rope breakage during service life. Therefore, in the design of the support system, the best solution is to ensure the structure's safety while considering rope breakage. Thus, the structure will not be destroyed immediately when any rope breaks.

3) Easy to adjust displacement. In some cases, acrylic shells may produce large structural displacements. The structural displacement of the acrylic shell is corrected by adjusting the length of the vertical rope. Therefore, the rope design must facilitate adjustment of the rope length.

4) Less contact area. The working efficiency of the JNE detector is related to the transparency of the acrylic shell. More neutrino information was captured by the PMTs when more Cherenkov radiation penetrated the surface of the acrylic shell. Therefore, minimizing the occluded area of the support system on the acrylic shell is essential to maximize transparency. Reasonable restraint methods can be selected, reducing the rope diameter while ensuring structural safety.

## 3. Structural scheme of the acrylic vessel

### 3.1 Structural Design

The JNE detector's acrylic shell will be installed in a chamfered rectangular stainless-steel frame measuring 14.5 m in length, 12.9 m in width, and 13.2 m in height. The detector core is an acrylic shell with a diameter of 9.96 m to achieve the volume requirement of 500 $m^3$. The acrylic vessel shell was preliminarily designed to be 50 mm, with an overall height of 14 m, weighing approximately 19 tons, and an internal capacity to hold about 500 $m^3$ of liquid.

Unlike SNO and JUNO detectors, which have liquid pipelines only at one end,

liquid pipes for both the upper and lower ends of the acrylic shell were specially designed for convenient fluid exchange. During operation, one end was used as the liquid inlet and the other as the outlet. This design is called 'chimney', with a diameter of 1 m and a thickness of 50 mm. The dimensions of the acrylic shell are shown in Fig. 4.

The acrylic shell was preliminarily designed to be supported by a system of 28 synthetic fiber ropes, with 14 ropes each in the top and bottom hemispheres, because the acrylic shell may be subjected to sinking or floating forces under various working conditions, as shown in Fig. 3. The suspension points of the ropes were aligned at both ends of the vessel. To determine a safe rope diameter, the reference working condition for the acrylic shell, labeled Working Condition 2, was used for the calculations. A diameter of 35 mm was selected for the synthetic fiber ropes, according to a preliminary estimation.

In the final structural design of the acrylic shell, as shown in Fig. 4, each hemisphere is supported by two layers of horizontal ropes, each sustained by seven vertical ropes. The two layers of rope in the hemisphere are independent of each other. This support method distributes the restraining force of the ropes across the horizontal ropes, resulting in lower stress on the surface of the acrylic vessel. More importantly, the failure of any rope within a hemisphere does not precipitate structural collapse. Each rope was equipped with a tensioning apparatus for length adjustment and pre-tensioning, aiding in the spatial calibration of the sphere. The different placement of the horizontal ropes induced varying stresses on the acrylic shell, further analyzed and optimized in Section 3.2.

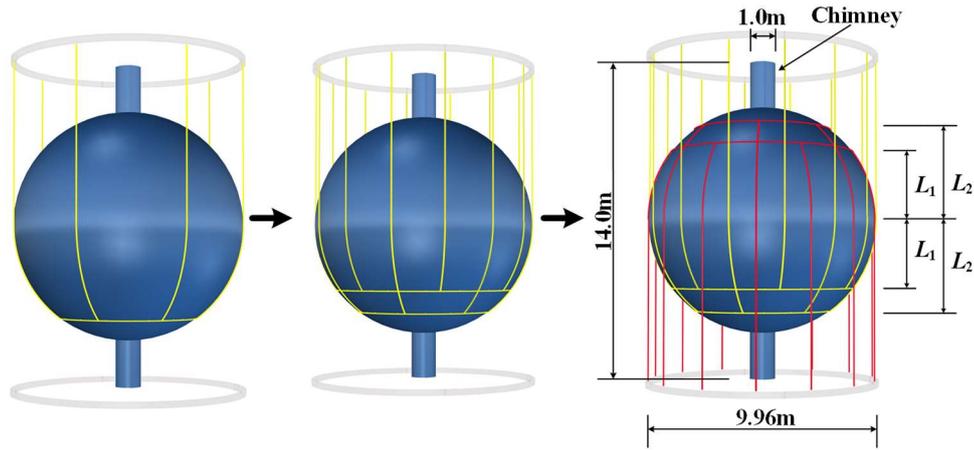

Fig. 4 The scheme of the acrylic shell for the Jinping Central Detector

## 3.2 Structural Optimization

In the structural scheme, the positions of the horizontal ropes were adjustable. Controlling the placement of the horizontal ropes and the distance between the two layers of nets in the same hemisphere is crucial for minimizing the maximum von Mises stress on the acrylic vessel. For the analysis, different scenarios were considered: the distance from the upper horizontal rope to the shell's equator, with $L_1$ being 2.5 m, 3 m, 3.5 m, and 4 m; the distance from the lower horizontal rope to the shell's equator, which writes it $L_2$ being 3.25 m, 3.5 m, 3.75 m, 4 m, 4.5 m, and 5.0m. The relationship between $L_1$ and $L_2$ is illustrated in Fig. 4 (a).

The FEA software ABAQUS was employed to perform mechanical analysis on the developed structural model. In the FEA model, the friction coefficient between the acrylic shell and ropes was set to 0.3, and a surface-to-surface contact was used. The material properties of the PMMA and synthetic fiber ropes are listed in Table 2. The materials defined in the FEA were all elastic because their stress was controlled at a low level, far from the plastic stage. The acrylic shell was simulated using C3D8R elements with a mesh density of 0.4 and localized refinement to 0.2; synthetic fiber ropes were modeled using B31 elements with a mesh density of 0.1. Working condition 2 in Table 1 was adopted herein.

**Table 2 Mechanical properties**

| Materials | Young's modulus (MPa) | Density (kg/m³) | Poisson ratio |
|---|---|---|---|
| PMMA | 3000 | 1190 | 0.376 |
| Synthetic fiber rope | 1500 | 1410 | 0.46 |

The calculated results are detailed in Table 3 and Fig. 5. It is found that when the distance $L_1$ is less than 2.5 meters, the horizontal ropes tend to slip as the nets are positioned closer to the equator, their radius increases, leading to significant deformation under load. Such deformation precipitates the acrylic shell falling out of the horizontal ropes. When the distance $L_2$ exceeds 4.5 meters, the contact area between the horizontal ropes and the acrylic shell becomes too small, diminishing the effective circumferential restraint on the acrylic shell. This could result in the acrylic vessel slipping out through the gaps in the vertical ropes. The distance between the two layers of rope can be expressed as $L_2$-$L_1$. As depicted in Fig. 5(a), when the spacing between two horizontal ropes on the same hemisphere is too tiny, the Mises stress intensifies due to the range of applied forces being too concentrated. As seen in Fig. 5(b), the greater the spacing between horizontal ropes, the higher the maximum force exerted on the ropes. This occurs because the total downward force acting on the vertical ropes is constant in a given scenario. When the spacing between two layers of ropes on the same hemisphere increases, the force distribution on the two layers becomes more unbalanced, with the layer near the equator carrying a smaller force and the layer away from the equator receiving a more significant force.

**Table 3 FEA results of the various models (Working Condition 2)**

| $L_1$ (m) | $L_2$ (m) | Maximum von Mises stress on acrylic (MPa) | Maximum principal stress on acrylic (MPa) | Maximum Axial force of ropes (kN) | Maximum structural displacement (mm) |
|---|---|---|---|---|---|
| 3.00 | 3.25 | 5.87 | 2.12 | 123.2 | 198 |
| 3.25 | 3.50 | 5.04 | 2.12 | 110.8 | 185 |
| 3.50 | 3.75 | 4.48 | 2.21 | 106.7 | 177 |
| 3.75 | 4.00 | 4.06 | 2.63 | 101.0 | 162 |
| 2.50 | 3.00 | | The rope loop slipped out | | |
| 3.00 | 3.50 | 4.10 | 2.15 | 139.7 | 208 |
| 3.50 | 4.00 | 2.73 | 2.15 | 108.9 | 174 |
| 4.00 | 4.50 | 2.49 | 2.30 | 101.2 | 154 |
| 2.50 | 3.50 | | The rope loop slipped out | | |
| 3.00 | 4.00 | 2.28 | 2.15 | 117.9 | 174 |
| 3.50 | 4.50 | 2.00 | 2.27 | 96.5 | 157 |
| 4.00 | 5.00 | | Insufficient constraint | | |

Considering the advantages and disadvantages, the horizontal rope configuration with $L_1 = 3$ m and $L_2 = 4$ m was the best option. Because in this scheme, the acrylic ball has a small stress, the axial force of the rope meets the requirements, and the placement of the rope is moderate. Although the maximum stress on the acrylic and the maximum axial force of the ropes are smaller when $L_1 = 3.5$ m and $L_2 = 4.5$ m, the placement of the rope is close to the insufficient constraint risk line shown in Fig. 5, and therefore this scheme is not chosen. The selected scheme is circled in Fig. 5. The maximum principal stress on the acrylic shell is only 2.15 MPa (Fig. 6 (a)), well below 3.5 MPa. The maximum axial force on the rope appears on the lower horizontal rope, and the full axial force is 117.9 kN (Fig. 6 (c)), which is 11.8% of the breaking force of the rope. The structural displacement of the optimized acrylic shell was 174 mm (Fig. 6 (d)), which was well below 600 mm.

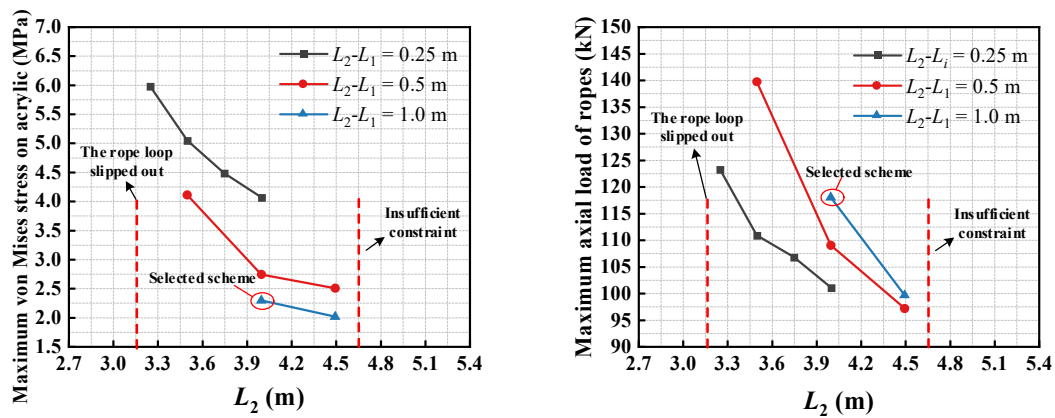

（a）Maximum Von Mises stress versus $L_2$ curve　　（b）Maximum rope load versus $L_2$ curve

**Fig. 5 Comparison of the results with various $L_1$ and $L_2$ values**

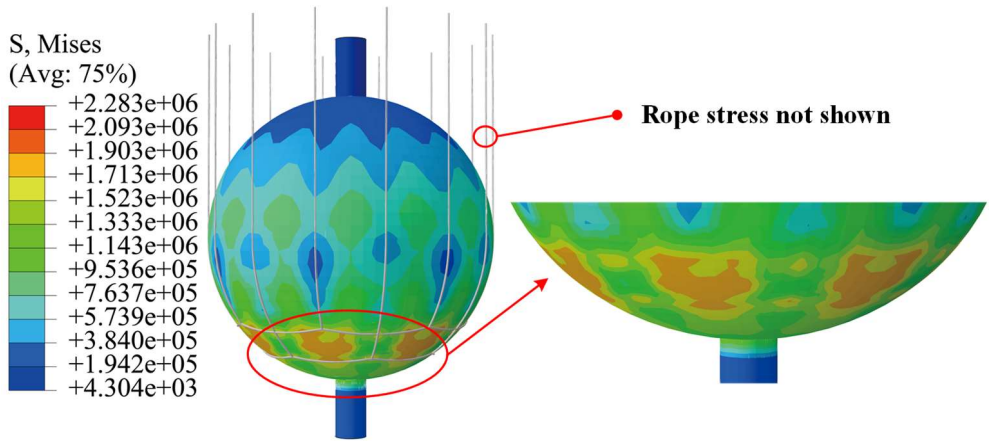

(a) Von Mises stress on acrylic (Pa) （Rope stress not shown）

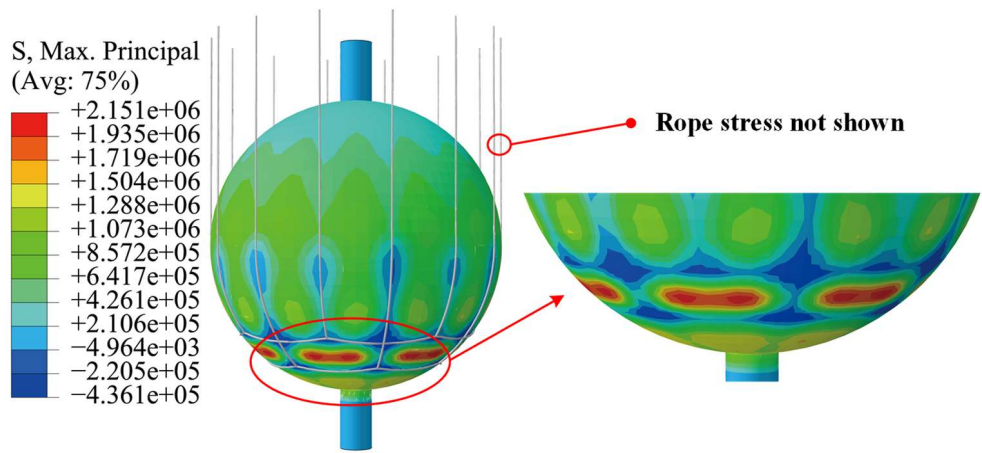

(b) Principal stress on acrylic (Pa) （Rope stress not shown）

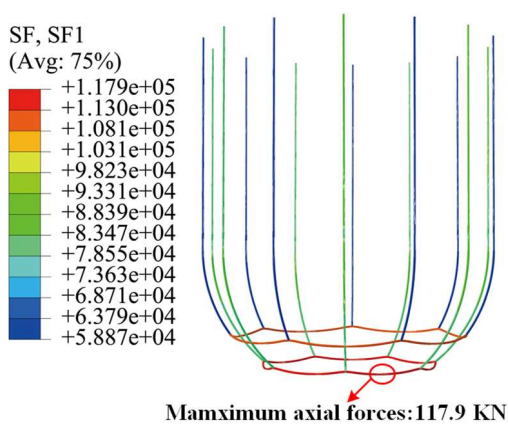
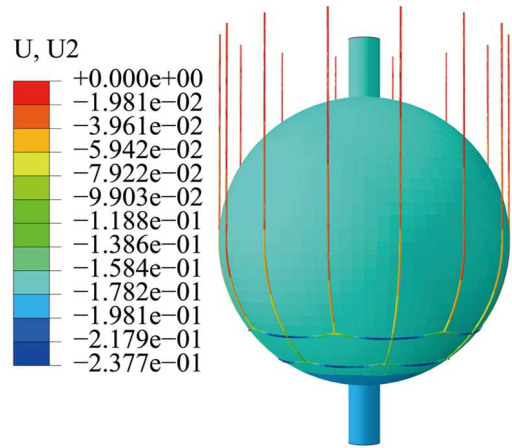

(c) Axial forces of rope (N)　　　　　　(d) Structural Displacement (m)

**Fig. 6 FEA results of optimized model (Working Condition 2)**

## 3.3 Structural Stability

The method for analyzing stability uses FEA software to conduct buckling calculations on the structure. This involves determining the critical buckling load of the structure and assessing the safety factor between the critical load and the maximum design stress. Working Condition 5 is adopted for the calculation.

Linear buckling analysis is first performed based on the assumption that PMMA is an ideal elastic material without flaws. However, PMMA is not a perfect material and may have manufacturing imperfections; therefore, a non-linear buckling analysis is subsequently performed. Because the von Mises stress on the sphere is low and the material remains elastic, this part of the analysis mainly addresses geometric nonlinearity.

Under Working Condition 5, the first buckling mode of the acrylic shell, calculated using ABAQUS, had a safety factor of 14.72. Fig. 7(a) shows the buckling form of the acrylic shell, with its lower segment particularly susceptible to brittle disruption. Fig. 7(b) presents the risk buckling load factor versus the deflection curve of some feature points on the shell by using the 'uniform imperfection method' with an imperfection amplitude set to 1/300 of the diameter of the shell. The peak of this curve indicates a nonlinear safety factor of 5.3, exceeding 4.0, confirming that the structural model meets all the mechanical performance criteria.

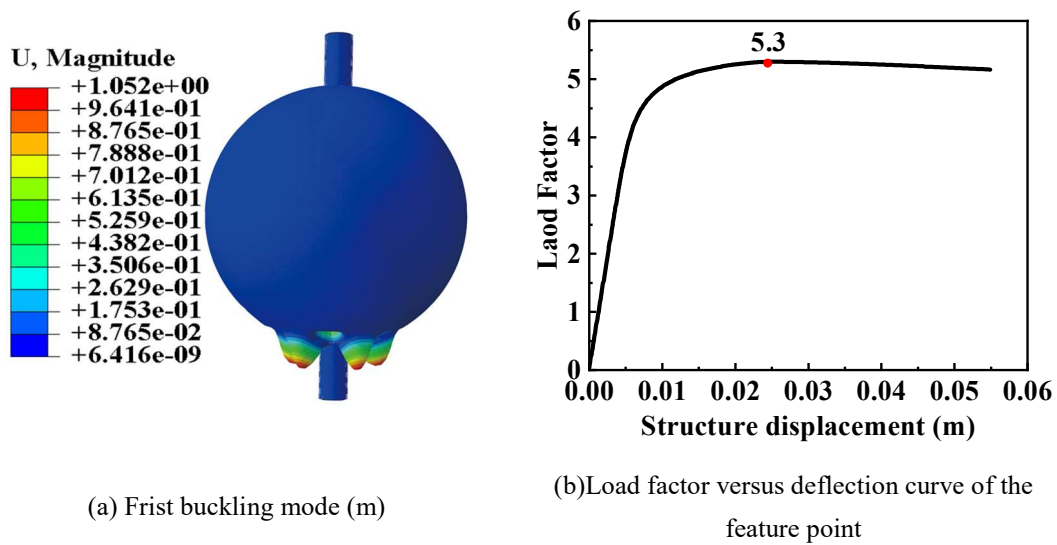

(a) Frist buckling mode (m)

(b) Load factor versus deflection curve of the feature point

**Fig. 7 Stability analysis for optimized model (Working Condition 5)**

## 3.4 FEA results under various working conditions

Section 2 mentioned that the JNE detector, designed to replace internal and external solutions easily, often faces various working conditions. This included conditions with different liquid levels and densities inside and outside the acrylic shell. These conditions, listed in Table 1, must be considered. To assess their impact on the structure, FEA is conducted for each condition. The results are compiled in Table 4.

Table 4 shows that the maximum stresses on the acrylic under long-term working conditions are lower than 3.5 MPa, and those under short-term working conditions are smaller than 7.0 MPa; the maximum axial forces of ropes are 122.6 kN, lower than 150 kN; the structural displacements are limited within 600 mm. It demonstrates that all the considered working conditions satisfy the design requirements.

Compared with Working Conditions 2, 3, and 4, the densities of the inner and outer liquids were the same, but the liquid levels were different. The maximum Mises stress in Condition 3 and 4 are greater than that in Condition 2 (Reference condition), which indicates that the appearance of the liquid level difference leads to an increase in Mises stress. The most disadvantageous condition - Condition 4 has the most significant first principal stress (see Fig. 8 (a)) of 2.99 MPa, a 39.1% increase compared to Condition 2 (see Fig. 6 (b)), where internal and external liquid densities are the same but without a level difference. Under Condition 4, the greater density of the internal liquid creates greater gravity than the floatage, and the liquid level difference amplifies this effect, increasing the principal stress on the acrylic shell.

Under Conditions 5, 6, and 7, the maximum stress of the sphere was concentrated in the upper hemisphere due to buoyancy, as shown in Fig. 8 (c) and (b). The von Mises stress in Conditions 6 and 7, with a liquid level difference, was more significant than in Conditions 5 (no liquid level difference). When the Condition 6 (high density with higher liquid level) occurs, the first principal stress shown in Fig. 8 (b) is 4.54 MPa, which is 288% higher than that under Condition 5.

The analysis showed that the von Mises stress increased when there was a liquid level difference between the inside and outside of the acrylic shell. In particular, when the level of a high-density liquid is higher than that of a low-density liquid, the gravity or buoyancy effect is amplified, increasing the first principal stress of the sphere.

Therefore, during the liquid-exchange process of the acrylic shell, the liquid levels inside and outside the sphere should be kept as equal as possible. In exceptional cases, the liquid level of low-density liquid should be higher than that of high-density liquid as much as possible.

**Table 4 FEA results for all working conditions**

| Working condition | Maximum von Mises stress on acrylic (MPa) | Maximum principal stress on acrylic (MPa) | Maximum axial force of ropes (kN) | Maximum structural displacement (mm) |
|---|---|---|---|---|
| 1 | 0.69 | 0.38 | 25.2 | 81 |
| 2 | 2.28 | 2.15 | 117.9 | 180 |
| 3 | 2.55 | 1.38 | 122.6 | 184 |
| 4 | 2.49 | 2.99 | 119.7 | 178 |
| 5 | 2.57 | 1.17 | 113.3 | 176 |
| 6 | 3.81 | 4.54 | 110.0 | 184 |
| 7 | 2.20 | 1.01 | 110.6 | 173 |
| 8 | 2.53 | 1.14 | 108.6 | 173 |
| 9 | 2.28 | 2.23 | 121.0 | 182 |

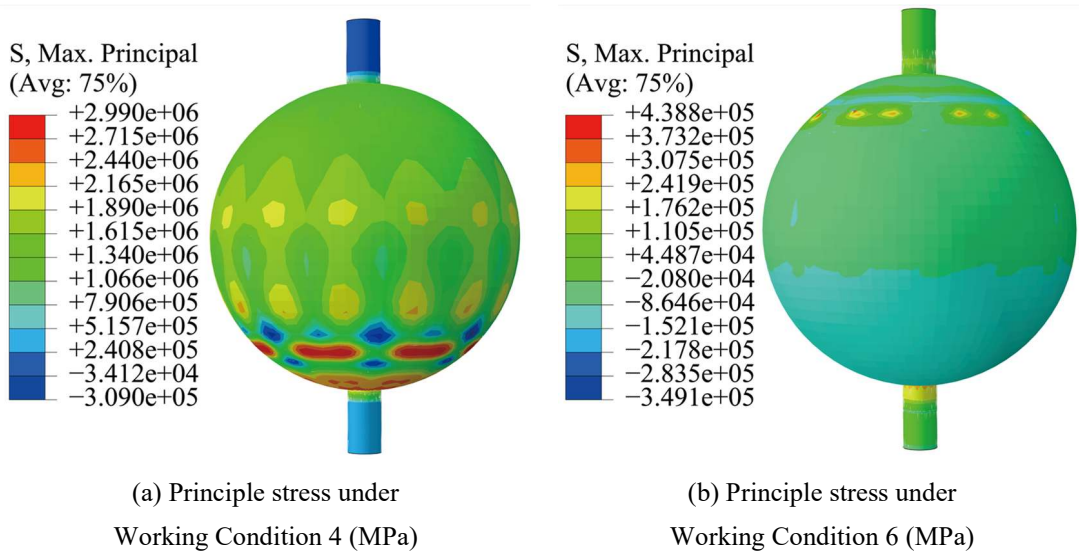

(a) Principle stress under Working Condition 4 (MPa)

(b) Principle stress under Working Condition 6 (MPa)

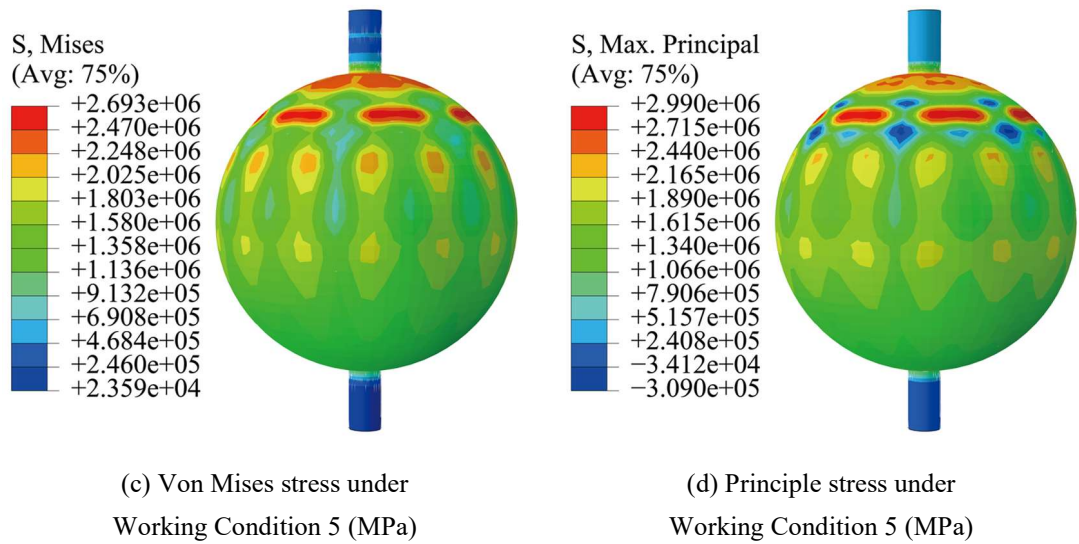

(c) Von Mises stress under
Working Condition 5 (MPa)

(d) Principle stress under
Working Condition 5 (MPa)

**Fig. 8 Stress on acrylic shell under Working Conditions 4, 5 and 6**

## 4. Analysis of other factors

### 4.2 Effect of temperature change

Located in the deepest underground laboratory in the world, the JNE detector is largely insulated by temperature variations. However, due to their high-temperature creep propensity, synthetic fiber ropes experience accelerated creep rates at elevated temperatures. Temperature variations can induce substantial elongation of the ropes, consequently causing a marked displacement of the acrylic shell. Additionally, acrylic shell and synthetic fiber ropes have different linear expansion coefficients. Therefore, temperature fluctuations may have caused stress changes in the acrylic shell. Hence, the potential impact of temperature change must be considered. The normal temperature in the Jinping laboratory is around 20°C, with maximum variations unlikely to exceed 10 °C. For safety, a temperature range of 0 to 40 °C is assumed. The linear expansion coefficient for the acrylic shell is taken as $8.0 \times 10^{-5}$ °C$^{-1}$, and for the synthetic fiber ropes, it is $2.0 \times 10^{-5}$ °C$^{-1}$. The FEA was conducted under Working Condition 2 using these parameters.

As per the results shown in Table 5, when the working environment temperature fluctuates between 0 °C and 40 °C, the stress on the acrylic shell consistently meets the requirement of not exceeding 3.5 MPa, and the maximum axial force of the rope is 122.7 kN, which is 12.3% of braking force. The impact of temperature on stress is minimal and is not a controlling factor in structural design.

High temperatures accelerate the creep process of suspension ropes. Therefore, although the effect of temperature variation on stress is minimal, maintaining a stable long-term environmental temperature is advisable.

**Table 5　Effect of environmental temperature (Working Condition 2)**

| Temperature (°C) | Maximum von Mises stress on acrylic (MPa) | Maximum principal stress on acrylic (MPa) | Axial force of ropes (kN) | Maximum structural displacement (mm) |
|---|---|---|---|---|
| 40 | 2.27 | 2.18 | 116.7 | 183 |
| 30 | 2.23 | 2.26 | 119.3 | 182 |
| 20 | 2.28 | 2.15 | 117.9 | 180 |
| 10 | 2.30 | 2.16 | 122.7 | 188 |
| 0 | 2.21 | 2.16 | 116.3 | 195 |

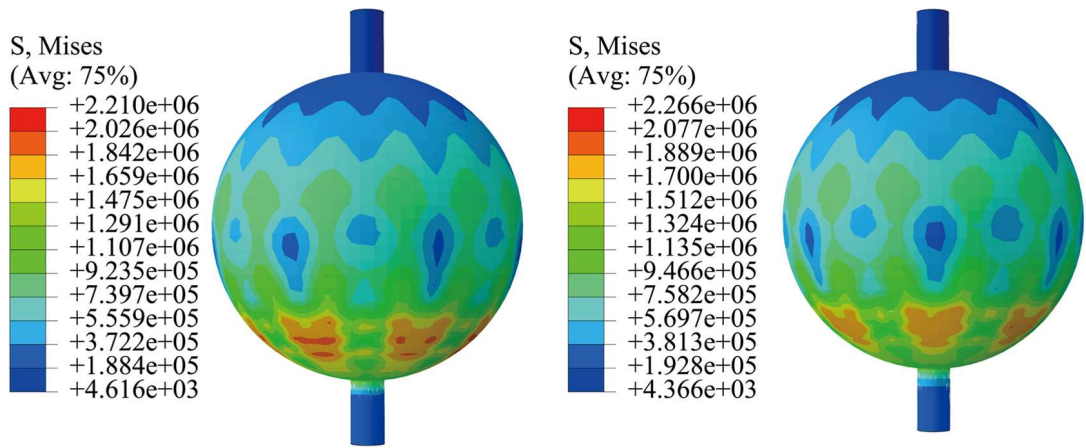

(a) Von Mises stress on acrylic at 0 ℃ (MPa)　　(b) Von Mises stress on acrylic at 40 ℃ (MPa)

**Fig. 9 FEA result at 0 ℃ and 40 ℃ (Working Condition 2)**

## 4.3 Effect of rope failure

Although the JNE detector has a long operational lifespan of 20 years, the synthetic fiber ropes suspending the acrylic shell are regularly unloaded and replaced at set intervals. Moreover, a considerable safety margin exists for the forces exerted on these ropes. This makes the likelihood of a rope breaking extremely low. Nevertheless, a detailed risk analysis is necessary to ensure that the acrylic shell support system remains intact even when some of the ropes fail.

In the event of rope failure, there are two distinct scenarios: 1)The vertical rope failure. 2) The horizontal rope fails. In the first scenario, where a vertical rope breaks, and the acrylic shell tilts and redistributes its load to the remaining 13 ropes. The pivotal aspect to observe was that the maximum displacement of the sphere in any direction should not exceed 600 mm. In the second scenario, where a horizontal rope fails, all seven vertical ropes connected to it simultaneously lose effectiveness. This transfers all the forces to the remaining seven ropes. While the sphere continues to experience balanced forces, the primary concern is ensuring that the stress on acrylic does not escalate. In Fig. 10, five envisioned scenarios of rope failure are shown. Failure modes 1, 2, and 3 demonstrate the breakage of vertical ropes, whereas failure modes 4 and 5 detail instances of horizontal rope failures.

Table 6 lists the results of the FEA calculations. When any rope suddenly fails, the Mises stress on the acrylic shell exceeds the long-term stress requirement of 3.5 MPa but does not exceed the short-term stress requirement of 7.0 MPa. In addition, the failure of any rope results in significant structural displacement. Specifically, failure mode 1 and failure mode 2 are both a vertical rope broken in a certain layer, but they produce greater Mises stress than failure mode 3. This is due to the force imbalance across the two layers of ropes in the same hemisphere after failure modes 1 and 2, leading to a majority of the force being borne by the layer without the failure, resulting in increased stress. In contrast, failure mode 3 facilitates a more balanced force redistribution on the acrylic shell due to concurrent breakages in both rope layers. But this mode causes significant tilting of the sphere, as shown in Fig. 11(d), with structural displacement reaching 491 mm, a 173% increase. Although this does not exceed the limit for structural displacement, it poses a considerable risk.

Failure modes 4 and 5, which involve the horizontal ropes' breakage, represent the least desirable scenarios. Table 6 shows that these modes result in considerable Mises stress, with 4.66 MPa for mode 4 and 5.09 MPa for mode 5, increases of 104% and 123%, respectively. The Mises stress for failure mode 5 is depicted in Fig. 11(a). When any horizontal rope on a hemisphere breaks, all seven ropes in that layer fail, leaving the acrylic shell supported only by an intact layer. This results in drastic stress redistribution and elevation. Fortunately, this type of failure does not result in an imbalance of the acrylic shell or breakage of the support cables; therefore, the resulting structural displacements are small, and the stress still meets the short-term requirements. The maximum axial force of the rope was 212.8 kN, 80.5% of the breaking force, shown in Fig. 11 (c).

Analytical insights reveal that while the likelihood of rope failure is minimal, it nonetheless poses a severe threat to the structural safety of the JNE detector acrylic shell. Such shortcomings may induce excessive stress, significant structural displacements, and rope forces, potentially leading to structural collisions. Consequently, it is necessary to regularly inspect and replace the ropes during use to avoid such risks. The analysis also proved that the support system, even in unique failure scenarios, can uphold the short-term structural safety of the sphere.

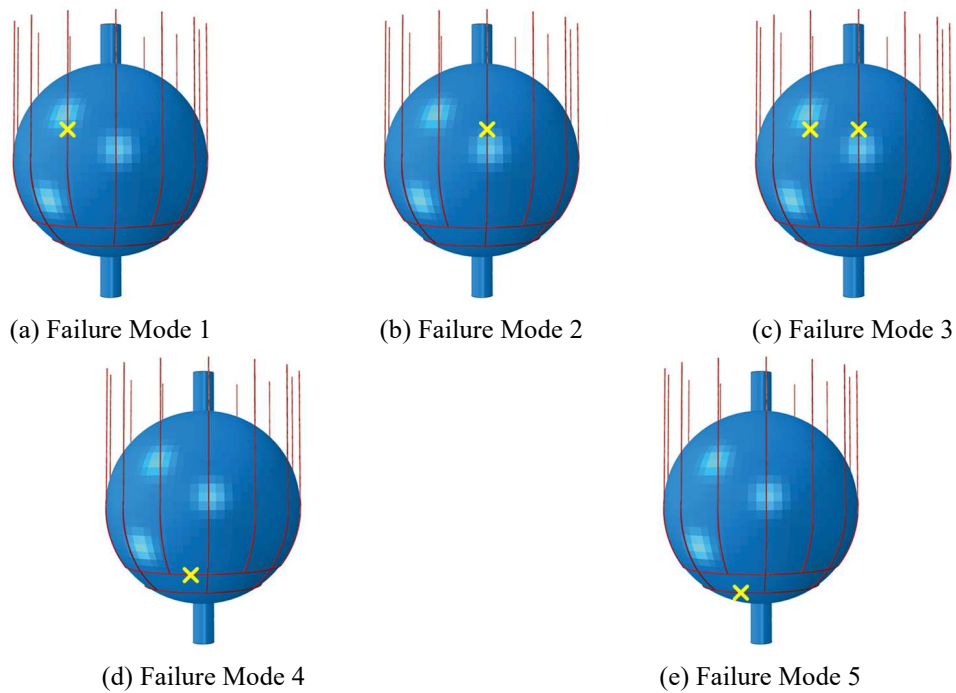

Fig. 10 Forms of rope failure

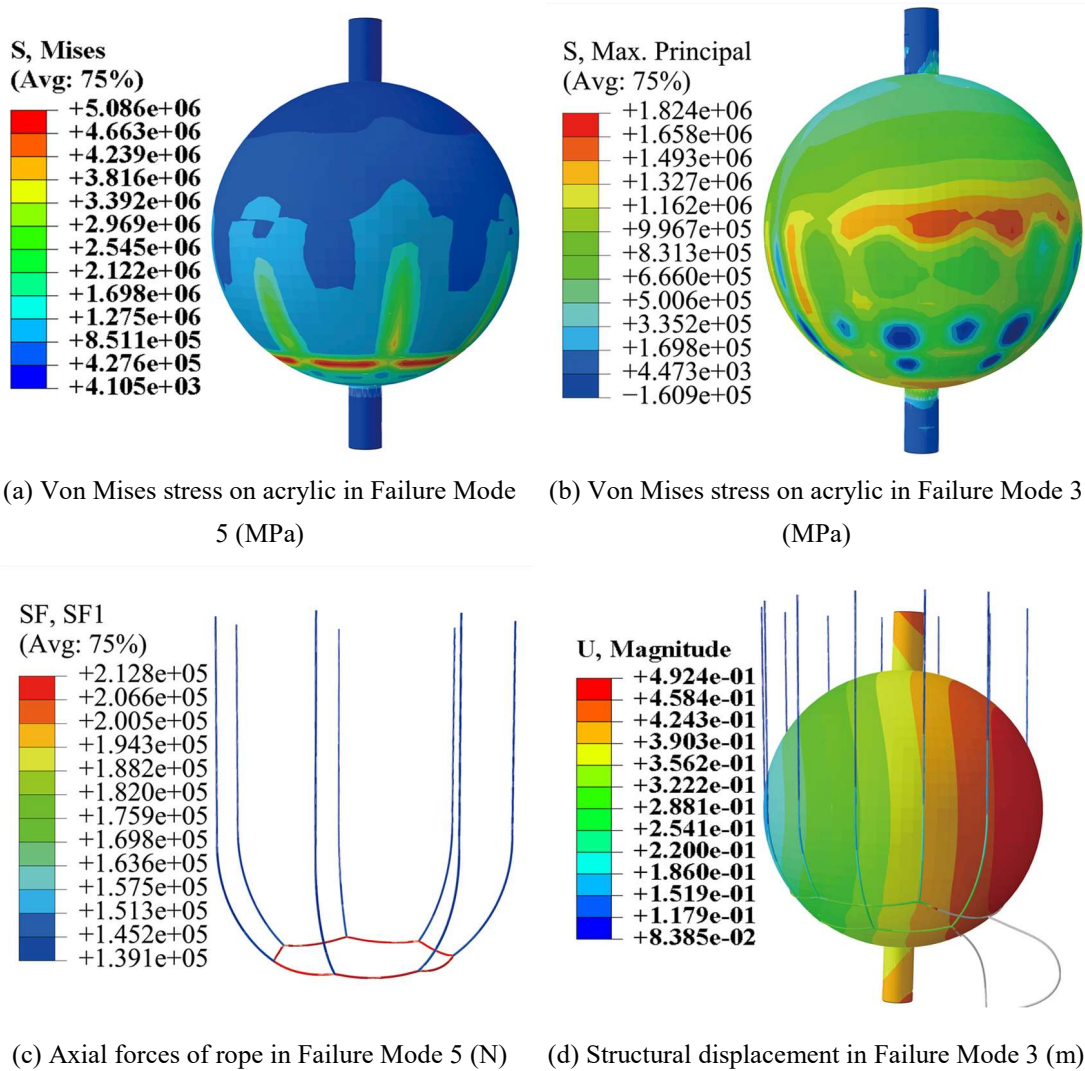

(a) Von Mises stress on acrylic in Failure Mode 5 (MPa)

(b) Von Mises stress on acrylic in Failure Mode 3 (MPa)

(c) Axial forces of rope in Failure Mode 5 (N)

(d) Structural displacement in Failure Mode 3 (m)

**Fig. 11 FEA results under rope failure**

**Table 6 Summary of FEA calculations under rope failures**

| Forms of rope failure | Maximum von Mises stress on acrylic (MPa) | Maximum principal stress on acrylic (MPa) | Axial force of the rope (kN) | Maximum structural displacement (mm) |
| --- | --- | --- | --- | --- |
| Mode 1 | 3.62 | 1.43 | 152.3 | 249 |
| Mode 2 | 4.20 | 1.54 | 166.2 | 276 |
| Mode 3 | 3.33 | 1.82 | 134.2 | 491 |
| Mode 4 | 4.66 | 1.47 | 193.2 | 263 |
| Mode 5 | 5.09 | 1.48 | 212.8 | 262 |

## 4.4 Effect of Young's modulus of ropes

Synthetic fiber ropes for the support system of the acrylic shell in the JNE detector are still being selected, with ongoing experimental studies on their mechanical properties. In the finite element model, Young's modulus of the ropes was set to 15 GPa. However, different synthetic fiber ropes have different Young's moduli. Analysis of ropes with different Young's moduli is necessary to select rope materials.

Adjusting Young's modulus of the ropes (Table 7) does not significantly affect the stress on the acrylic shell; however, it substantially impacts the structural displacement. The results indicate that the lower the Young's modulus of the ropes, the greater the structural displacement of the acrylic shell. It is recommended that Young's modulus of the ropes should be greater than 3 GPa to ensure that the displacement does not exceed the 600 mm limit.

Table 7 Effect of Young's modulus of ropes (Working Condition 2)

| Young's modulus of ropes (GPa) | Maximum von Mises stress on acrylic (MPa) | Maximum principal stress on acrylic (MPa) | Axial force of ropes (kN) | Maximum structural displacement (mm) |
| --- | --- | --- | --- | --- |
| 3 | 2.50 | 2.32 | 107.2 | 560 |
| 5 | 2.43 | 2.29 | 106.3 | 366 |
| 10 | 2.34 | 2.22 | 116.9 | 229 |
| 15 | 2.28 | 2.15 | 117.9 | 180 |
| 20 | 2.32 | 2.28 | 124.7 | 155 |
| 25 | 2.33 | 2.36 | 126.5 | 139 |

## 5. Conclusion

The JNE detector benefits from its unique experimental environment and is a significant mission for exploring solar and other neutrinos. Consequently, the design of the core component—the acrylic shell—requires meticulous attention. This study presents a design for the acrylic shell of the JNE detector using Abaqus software, and the bearing capacity under various working conditions were performed, as well as the stability analysis. Additionally, the analyses of temperature effects, rope failure, and Young's modulus of the ropes are conducted, leading to the following conclusions:

(a) An acrylic shell for the JNE detector and optimized support scheme is designed. The final scheme consists of two horizontal rope layers per hemisphere strategically placed at distances of 3 m and 4 m from the equator. the maximum stresses on the acrylic under long-term working conditions are lower than 3.5 MPa, and those under short-term working conditions are smaller than 7.0 MPa; the maximum axial forces of ropes are 122.6 kN, lower than 150 kN; the structural displacements are limited within 600 mm. Through non-linear buckling analysis, the safety factor is found to be 5.3 in Working Condition 5, exceeding 4.0. This scheme meets the design requirements.

(b) The temperature effect analysis of the acrylic shell indicates that temperature is not a controlling factor in structural design, with the maximum Mises stress being only 2.3 MPa, satisfying the requirement.

(c) When ropes failure occurs, the maximum rope force was 21.2% of the rope-breaking force. The breakage of a single horizontal rope results in a maximum Mises stress of 5.09 MPa on the sphere, exceeding the 3.5 MPa limit but well below 7.0 MPa. The simultaneous breakage of the two vertical ropes causes a maximum structural displacement of 493 mm on the acrylic shell, below the 600 mm threshold. This structure remained safe and reliable, even in rope failure scenarios.

(d) The lower the Young's modulus of the ropes, the greater the structural displacement. When the Young's modulus of rope is less than 3 GPa, the displacement of the structure reaches 560 mm, approaching 600 mm. Young's modulus of the ropes is recommended to be greater than 3 GPa.

## Acknowledgments

This work was supported by the National Natural Science Foundation of China (Nos. 52208169 and 12127808), the Ministry of Science and Technology of China (No. 2022YFA1604704) and "the Fundamental Research Funds for the Central Universities" HUST (No. 2020kfyXJJS126).## References

[1]  M. Sajjad Athar, Steven W. Barwick, Thomas Brunner, et al., Status and perspectives of